\documentclass[12pt,a4paper]{article}%
\usepackage[utf8]{inputenc}
\usepackage{epsfig}
\usepackage{float}
\usepackage{graphicx}
\usepackage{hyperref}
\usepackage{color}
\usepackage{subfig}
\usepackage{amsmath}
\usepackage{amssymb}

\graphicspath{{./}}
\def\knn{k_{\rm nn}}

\begin{document}
\section{Multiplex Modeling of the Society}

\begin{footnotesize}
{\linespread{0.5}
{\bf J\'anos Kert\'esz${\bf^{1,2,3}}$, J\'anos T\"or\"ok${\bf^{1,2}}$, Yohsuke Muraze${\bf^{4,5}}$, Hang-Hyun Jo${\bf^{6,3}}$ and Kimmo Kaski${\bf^{3}}$}\\ \\
$^1$CEU, Center for Network Science, N\'ador u. 9, Budapest, H-1051, Hungary\\
$^2$BME, Institute of Physics, Budafoki \'ut 8, H-1111, Hungary\\
$^3$Department of Computer Science, Aalto University, P.O. Box 15500, Espoo, Finland\\
$^4$RIKEN Advanced Institute for Computational Science, Kobe, Hyogo 650-0047, Japan\\
$^5$CREST, Japan Science and Technology Agency, Kawaguchi, Saitama 332-0012, Japan\\
$^6$Pohang University of Science and Technology, Pohang 37673, Republic of Korea
}
\end{footnotesize}

\subsection{Introduction}

Networks of social interactions are paradigmatic examples for multiplexity. It was recognized long ago by social scientists~\cite{Fienberg1985Statistical,Wasserman1994Social} that the best way to interpret the network of different kinds of human relationships is a multiplex network, where each layer corresponds to a particular type of relationship, e.g., between kins, friends, or co-workers (see~\cite{Brodka2014Multilayered} and references therein). 

Until recently only rather small size networks could be studied due to limited size datasets 
collected by traditional methods of sociology~\cite{Bernard2000Social}. Consequently such a fundamental question, like the structure of the network of interactions at the societal level, could hardly be approached. In fact, the global consequences of local rules like formulated in the famous Granovetter hypothesis~\cite{Granovetter1973Strength} about the strength of weak ties could not be tested by the traditional 
methods. Over the past fifteen years this situation has changed substantially due to large scale of human sociality related datasets becoming increasingly available.
  
Social interaction between people can always be considered as a kind of communication. In the digital era much of the communication has shifted to channels of information-communication technology (ICT), where records are created about all interactions. Mobile phone calls, text messages, Social Network Services (SNSs) like Facebook and Twitter, and even massively multiplayer online games produce a deluge of data, which can be considered as digital footprints of individuals and thus serving as a gold mine for research of human sociality. Thanks to this development, a new discipline has emerged: Computational Social Science~\cite{Lazer2009Computational}.

Call detail records (CDRs) of mobile phones play a special role among datasets from today's communication tools~\cite{Blondel2015Survey} as the coverage is close to 100\% in the developed countries and most people make no step without their devices. The CDRs completed with metadata like gender, age, zip code, and information about location open up further research possibilities. Such data were used among others to prove the Granovetter hypothesis~\cite{Onnela2007Structure}, uncover regularities in human mobility patterns~\cite{Gonzalez2008Understanding}, and deduce the distance dependence of social ties~\cite{Lambiotte2008Geographical}. Using the metadata, it was also possible to distinguish between different types of relations and relate the activities to age and gender of the individuals~\cite{Palchykov2012Sex, Jo2014Spatial}. A large amount of observations have accumulated reflecting various interesting features of human interactions at the societal level~\cite{Onnela2007Analysis, Blondel2015Survey}. Many findings in the CDR dataset were found also to be characteristic to other communication channels,  e.g., emails~\cite{Kossinets2006Empirical}, Facebook~\cite{Lewis2008Tastes}, and Twitter~\cite{Myers2014Information}. Such features include broad distributions of network quantities like the degree and weight (to be defined later), community structure, and assortative mixing. This way a set of stylized facts have emerged~\cite{Murase2015Modeling}, and they serve as guidelines 
for large-scale modeling of the society. 

\begin{figure}[t]
\begin{center}
 \includegraphics[width=.99\columnwidth]{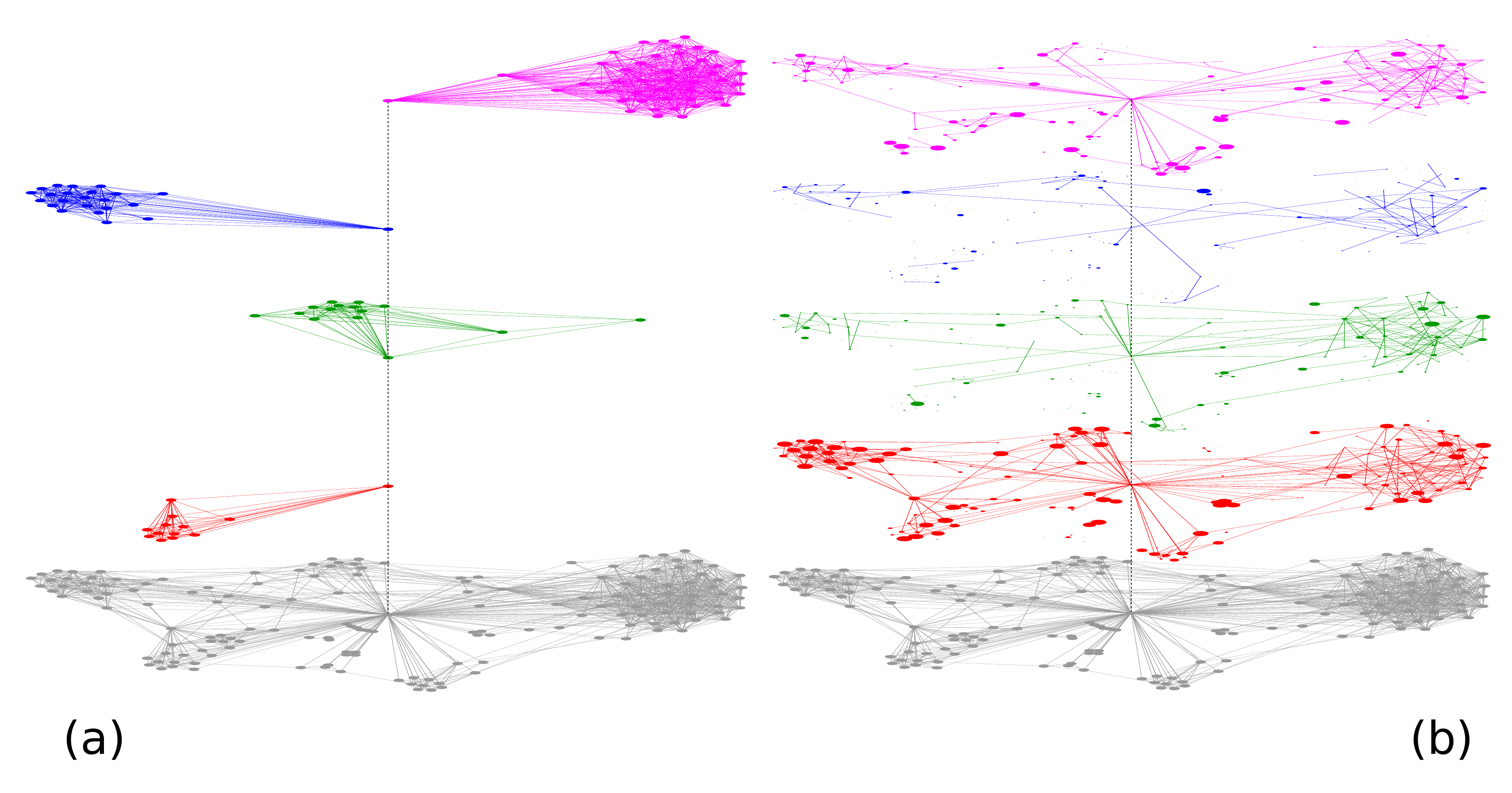}
 \caption{\small Egocentric network of a person (central vertical lines in the figure) as taken from the iWiW dataset, which, for simplicity, we assume to map out completely the person's relationships. This social network is resolved in two multiplex networks: (a) The layers correspond to different types of relationships or contexts as obtained from combination of metadata and community detection. Only the four most important relationships are shown (4 layers from the top). In (b) the layers represent communication channels and the top four of them are shown. In both cases the identical bottom layer is the aggregate or projection of the multiplex and contains all contacts.}
\label{Fig:Fig1}
\end{center}
\end{figure}

The society can be considered as a multiplex not only with respect to the different types of relationships but also from the point of view of 
the channels used for communication like face-to-face, mobile phones, and SNSs. For the latter case the layers of the multiplex correspond to the different communication channels. Figure~\ref{Fig:Fig1} illustrates these two different ways of considering multiplexity. 

A true picture about the entirety of human communication in the society should be based on comprehensive data from all the levels of this second type of multiplex. However, this is not feasible because even in the digital era not all forms of communication are registered. Moreover, data is usually available only for one channel, meaning that from the whole multiplex there is only one layer at our disposal for investigation. Linking data from diverse channels would of course be desirable but it is in most cases impossible due to the different origins of the data and/or for privacy reasons. 

Here we will discuss aspects of multiplexity in modeling the society. 
This chapter is organized as follows: First we sum up the ``stylized facts'' as obtained from so-called Big Data. Then we show how the Granovetterian structure, 
identified in single-layer data, can be modeled in a multiplex setup and how this structure can coexist with overlapping communities as they naturally emerge. In the next Subsection we report on modeling channel selection to analyze the sampling bias as introduced by single channel data. Finally we discuss the results and make an outlook. 

\subsection{Stylized facts for social networks}
\label{Sec:stylized}

In recent years, the availability of a large number of digital datasets have enabled us to characterize the structure of social networks in more detail and up to an unprecedented scale. For example, researchers have investigated emails~\cite{Eckmann2004Entropy, Kossinets2006Empirical}, mobile phone calls~\cite{Blondel2015Survey, Onnela2007Structure},  datasets from SNSs like Facebook~\cite{Ugander2011Anatomy} and Twitter~\cite{Kwak2010What} and even data of face-to-face proximity~\cite{Zhao2011Social, Isella2011Whats}. Analyses of such datasets revealed several commonly observed features or \emph{stylized facts} for social networks, as summarized in Table~\ref{table:summary}. Here we will mostly rely on the empirical findings from large-scale mobile phone call datasets~\cite{Onnela2007Analysis} because, due to the large coverage, they are expected to 
reflect the features of real social networks to large extent.

The most apparent stylized fact is the broadness of the distributions of the network quantities, like the 
degree $k$, link weight $w$, and node strength $s$~\cite{Albert2002Statistical, Onnela2007Analysis}. The weight of a link quantifies the interaction activity between two nodes. The strength, defined as the sum of weights of links involving the node, typically quantifies the activity of that node. The distributions of these quantities, i.e., $P(k)$, $P(w)$, and $P(s)$, have been found not only broad but also overall decreasing, implying that individual and interaction activities are heterogeneous and the maximum of the distribution is at $k\approx1$. The latter is clearly not consistent with our common sense that in a society it is hard to find a person with only one or a few relationships. This discrepancy should be attributed to the sampling effects, which will be discussed later in this Chapter. The overall decreasing $P(w)$ can be interpreted as the prevalence of weak links or weak ties in social networks.

Homophily is one of the main organizing principles of tie formation in social networks~\cite{McPherson2001Birds} as people tend to get along with those, who have similar characteristics. Here we are interested in the structure of social networks thus we focus on the degree-degree correlation. This correlation has been quantified in terms of assortativity, which can be measured by the Pearson correlation coefficient between degrees of neighboring nodes~\cite{Newman2002Assortative}. Many social networks are found to be assortative. A simple way to detect assortativity is to measure the average degree of neighbors for nodes with degree $k$, denoted by $k_{\rm nn}(k)$. An increasing trend means assortativity, as found, e.g., for the CDR dataset ~\cite{Onnela2007Analysis}

High clustering is evident in social networks as explained by the saying ``friends of friends get easily friends''. It means that if $B$ and $C$ are both connected to $A$, there is high chance that they are also connected to each other. The local clustering coefficient of a node is measured as the number of links between its neighbors divided by the maximal possible number of such links. The average local clustering coefficient for nodes with degree $k$, denoted by $c(k)$, is found to be generally a decreasing function of $k$, e.g., see~\cite{Onnela2007Analysis}.

\begin{table*}
  \caption{\small
      Stylized facts in the CDR dataset compared to the expected behavior for the entire social network, adopted from~\cite{Murase2015Modeling}. The arrows indicate the general trend of the profile. $\nearrow$ ($\searrow$) implies that the profile is monotonically increasing (decreasing). The initially increasing and then decreasing behavior is denoted by $\nearrow \searrow$. The definitions of quantities are described in the main text.
    \label{table:summary}
  }
\center
  \begin{tabular}{c|c|c}
      \hline 
    & CDR & Expected behavior\\ \hline
    $P(k)$ & $\searrow$ & $\nearrow \searrow$ \\
    $P(s)$ & $\searrow$ & $\nearrow \searrow$ \\
    $P(w)$ & $\searrow$ & $\searrow$ \\
    $s(k)$ & $\nearrow$ & $\nearrow$ \\
    $k_{\rm nn}(k)$ & $\nearrow$ & $\nearrow$ \\
    $O(w)$ & $\nearrow \searrow$ & $\nearrow$ \\
    $c(k)$ & $\searrow$ & $\searrow$ \\
    \hline
  \end{tabular}
\end{table*}

How individuals distribute their limited resources like time among their neighbors is also indicative to characterize the social networks. For this, the egocentric network, consisting of a node and its neighbors, has been studied in terms of the ranks of link activities or weights. A layered structure in the activity-rank relation has been claimed~\cite{Dunbar2011Constraints}, while smooth functional forms have been seen to fit with the 
empirical observations on the single channel data~\cite{Song2013Connections, Saramaki2014Persistence}.

Finally, on the mesoscopic scale of social networks we find a rich community structure. It means that nodes in communities are densely connected, while nodes between different communities are sparsely connected~\cite{Fortunato2010Community}. This picture is important to account for large clustering in sparse social networks with inhomogenous degree distribution where high degree nodes or hubs occur. Such topological property is correlated with the activities of links in that the communities of strongly connected nodes are weakly connected to each other~\cite{Onnela2007Structure}, in agreement with the famous Granovetter's hypothesis~\cite{Granovetter1973Strength}. Link-level consequences of weight-topology correlation can be measured by the average overlap for links with weight $w$, denoted by $O(w)$. The overlap of a link is the number of common neighbors of nodes connected by the link divided by the total number of neighbors of those nodes. It has been found that the stronger links show larger overlap~\cite{Onnela2007Structure} up to 95\% of the weights, thus showing agreement with the Granovetter's hypothesis.

It should be emphasized that these stylized facts have been deduced from single-layer data, representing one layer of the multiplex in Fig.~\ref{Fig:Fig1}(b). One such layer may reflect some multiplex properties stemming from different types of relationships as depicted in Fig.~\ref{Fig:Fig1}(a), while this restriction introduces some bias as we will show later in this Chapter. 

\subsection{Weighted multilayer model}

In order to reproduce the stylized facts shown in the previous section, a simple model was proposed by Kumpula~\emph{et al.}~\cite{Kumpula2007Emergence}, which we will call Weighted Social Network (WSN) model. This model succeeded in reproducing various stylized facts including community structure, Granovetterian weight-topology relation, assortative mixing, decreasing clustering spectrum, and relationship between node strength and degree. However, the WSN model has only a single-layer thus important aspects of the multilayer structure of social networks are missing. 

As discussed in the Introduction, people are involved in different social contexts or relationships and their social network should strongly depend on the context~\cite{Jo2012Spatiotemporal, jo2013contextual}. To handle these aspects, the social networks must be represented as a multilayer network or a multiplex~\cite{Kivela2014multilayer, boccaletti2014structure,jo2006immunization}, where links in the different layers correspond to different contexts, see Fig.\ref{Fig:Fig1}(a). These contexts are hardly distinguishable from the available data thus observed networks should be considered as an aggregate of the multiple layers. It is therefore a challenge to construct a model that 
reflects the observations and, at the same time,  
has the multilayer structure. In the following we discuss the possibilities of generalizing the WSN model~\cite{murase2014multilayer}, and show the conditions to reproduce the combination of Granovetterian weight-topology relationship and the overlapping communities arising from the multiplex nature of social networks.

\begin{figure}
\begin{center}
  \subfloat[]{
\includegraphics[width=.4\textwidth]{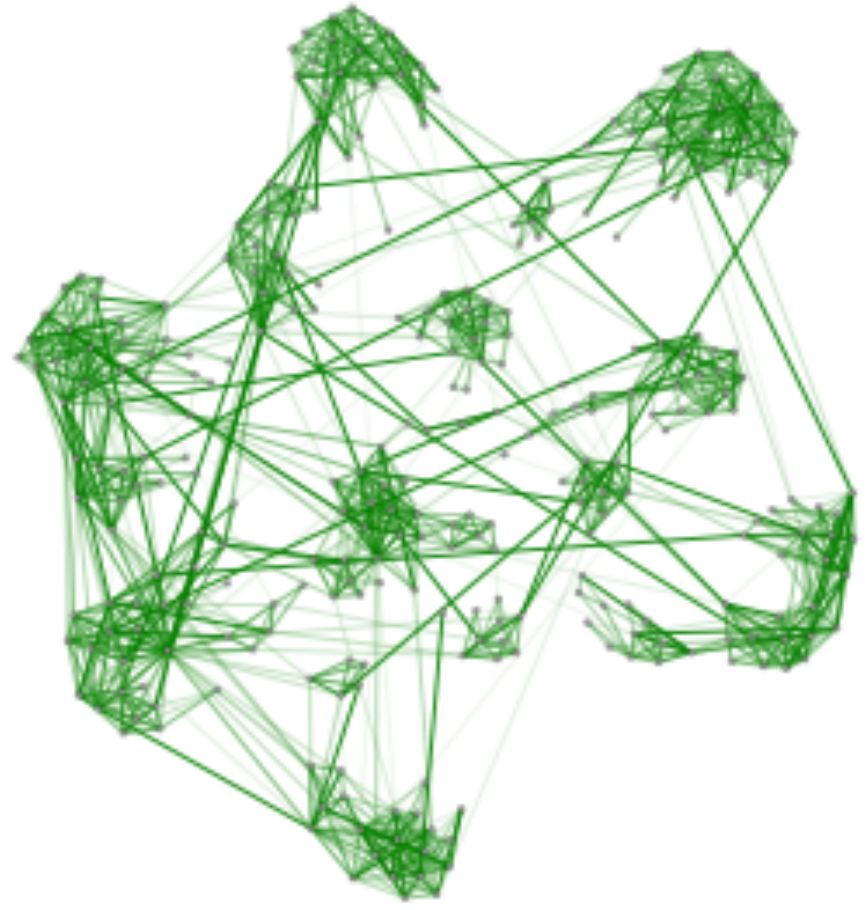}
  \label{fig:single_layer_snapshot}
}
  \subfloat[]{
\includegraphics[width=.45\textwidth]{alpha6_snapshot}
  \label{fig:geographic_snapshot}
}
\caption{\small
  Network snapshots for (a) the single-layer WSN model and (b) the geographic multilayer WSN model for $\alpha=6$.
  In (b), links in the first and the second layers are shown in thin black lines and thicker gray lines, respectively.
}
\label{fig:multilayer_snapshots}
\end{center}
\end{figure}

Let us first summarize the original WSN model~\cite{Kumpula2007Emergence}. It considers an undirected weighted network of $N$ nodes. The links in the network are updated by the following three rules.  The first rule is called {\it local attachment} (LA). Node $i$ chooses one of its neighbors $j$ with probability proportional to $w_{ij}$, which stands for the weight of the link between nodes $i$ and $j$. Then, node $j$ chooses one of its neighbors except $i$, say $k$, randomly with probability proportional to $w_{jk}$. If node $i$ and $k$ are not connected, they get connected with probability $p_{\Delta}$ by a link of weight $w_{0}$; if they have already been connected the weights of the links in the $(ijk)$ triangle, namely $w_{ij}$, $w_{jk}$ and $w_{ik}$ are increased by $\delta$. The second rule is {\it global attachment} (GA), where a node is connected to a randomly chosen node with weight $w_{0}$. This happens with probability $1$ if the node has no links, otherwise with probability $p_r$. Finally, the third rule, {\it node deletion} (ND) is introduced to the model, where with probability $p_{d}$, a node loses all its links. LA, GA, and ND are applied to all nodes at each time step, and we obtain a statistically stationary state after a sufficient number of updates. A snapshot of a network generated by this model is shown in Fig.~\ref{fig:multilayer_snapshots}(a).

It is clear by visual inspection that the single-layer WSN model does not generate significant amount of overlapping communities. This is a consequence of the LA rule. Even if one node happens by chance to belong to two communities, such communities tend to be connected by the links created with LA including the bridging node. While LA mechanism is crucial for generating community structure, it tends to merge communities. Thus, a mechanism to keep overlapping communities being separated must be incorporated to reproduce overlapping communities found in reality. One simple and plausible way of modeling this is the introduction of the multilayer structure, as this is the main cause of the overlapping communities.

\subsubsection{Uncorrelated multilayer WSN model}\label{sec:uncorrelated_multilayer}

In order to study multilayer effects, we first generalize the single-layer WSN model in a naive way as follows.  We consider $L$ layers of the same set of nodes and we assume that each layer corresponds to a different type of relationship or communication context.  For each layer, we independently construct a network in the same way as in the original single-layer WSN model.  After the networks are constructed in each layer, the aggregate network is created by summing up the edge weights: $w_{ij}=\sum_{k=1}^L w_{ij}^k$, where $w_{ij}^k$ is the weight of the link between nodes $i$ and $j$ in the $k$-th layer. It is this aggregate network for which we expect the coexistence between the overlapping community structure and the stylized facts already reproduced by the original WSN model.
In the following, $N = 50000$, $p_r = 0.0005$, $p_{\Delta} = 0.05$, $p_d =0.001$, $\delta=1$, and $w_0 =1$ are used. The results are obtained after $25 \times 10^3$ time steps and averaged over 50 realizations.

\begin{figure}[t]
\begin{center}
\includegraphics[width=.75\textwidth]{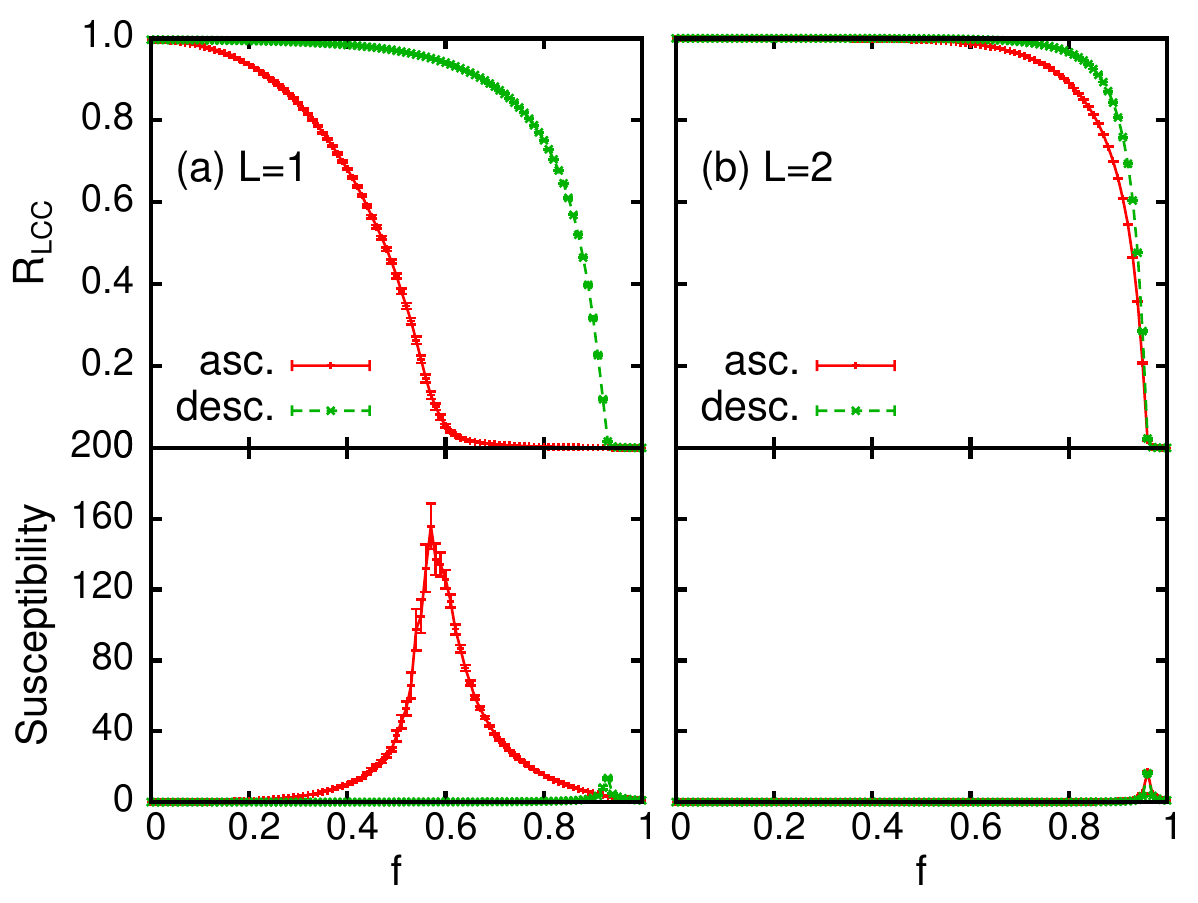}
\caption{\small
Link percolation analysis for $L=1$ (left) and $L=2$ (right). The upper panels show the relative size of the largest connected component, $R_{LCC}$, as a function of the fraction of the removed links $f$. The lower panels show the susceptibility $\chi$. Solid (dashed) lines correspond to the case when links are removed in ascending (descending) order of the link weights. The figure was taken from~\cite{murase2014multilayer}.
}
\label{fig:percolation_L1L2}
\end{center}
\end{figure}

It turns out that this naive multilayer model does not fullfill the expectations. Figure~\ref{fig:percolation_L1L2} shows the percolation analysis for a single-layer network ($L=1$) and a double-layer network ($L=2$) to verify the existence of the Granovetterian structure.  These two plots show the results for link removal in ascending and descending orders of the link weights. We define $f_c^a$ ($f_c^d$) as the percolation threshold for ascending (descending) order, marked by the disappearance of the largest connected component and the peak in the second moment of the component size distribution (also called susceptibility). The Granovetterian structure is characterized by a significantly large value of the difference $\Delta f_c=f_c^d - f_c^a$ between the two threshold values, as for the descending order the network gets earlier fragmented.

For $L=1$ we get $\Delta f_c \approx 0.35$, while for $L=2$ the percolation threshold for ascending order $f_c^a$ is approximately the same as that for descending order $f_c^d$, leading to $\Delta f_c \approx 0$. This indicates that the introduction of the second layer destroys the Granovetterian structure. The percolation threshold agrees well with that of an Erd\H{o}s-R\'enyi (ER) random network having the same average degree $\langle k \rangle$ as the simulated model: $f_c = 1 - 1/\langle k \rangle$ with the measured $\langle k \rangle = 21.9$. This observation shows that combining already two independent layers of the original single-layer WSN model leads to a high level of randomization in the aggregate model. Since strong links, which are intra-community links in the layer one, bridge the communities in the second layer, the difference between the roles of the links with different strength of weights disappears. This simulation results indicate that the empirical networks in different communication contexts cannot be independent. Hence inter-layer correlations play a pivotal role when modeling the multiplex structure of the social network.

\subsubsection{Geographic multilayer WSN model}\label{sec:geographic_multilayer}

The above results show that correlations between layers are essential in order to have $\Delta f_c$ for a multilayer model significantly different from zero, i.e., to reproduce the Granovetterian structure in a multiplex setting.
Previous studies have reported that there are strong geographic constraints on social network groups even in the era of the Internet~\cite{onnela2011geographic} and this is reflected in the CDR data~\cite{Krings2009Urban, Lambiotte2008Geographical, expert2011uncovering}.
For example, intercity communication intensity is inversely
proportional to the square of their Euclidean distance, which is
reminiscent of the gravity law~\cite{Krings2009Urban,
Lambiotte2008Geographical,Lengyel2015Geographies}. 

Motivated by these observations, we consider a model embedded into a
two-dimensional geographic space. Nodes are given fixed position in
the unit square with periodic boundary condition, which is shared by
all layers. The probability of new links created by the global
attachment (GA) process is proportional to $r_{ij}^{-\alpha}$, where
$r_{ij}$ is a distance between nodes $i$ and $j$, where $\alpha$ is a
new parameter controlling the dependence on geographic distance as
in~\cite{Kosmidis2008Structural,daqing2011dimension}. When $\alpha=0$,
this probability is independent of the geographic distance, thus the
model is equivalent to the uncorrelated multilayer model we presented
in the previous Subsection. When $\alpha$ is larger, the nodes will have
tendency to be connected with nodes that are geographically closer. Since GA
process creates links between non-connected nodes we choose the
following normalized connection probabilities in GA:
\begin{equation}\label{eq:alphadef}
p_{ij} = \frac{ r_{ij}^{-\alpha} }{\sum_{k \in S_i} r_{ik}^{-\alpha} },
\end{equation}
where $S_i$ is the set of the nodes not connected to the node $i$.
The other rules such as LA or ND are kept the same as in the original
WSN model.  Because the network for larger $\alpha$ has a smaller
average degree, we used a larger value of $p_r=0.002$, in the
following in order to keep the average degree comparable with the
results for the non-geographic model.

\begin{figure}
\begin{center}
\includegraphics[width=.7\textwidth]{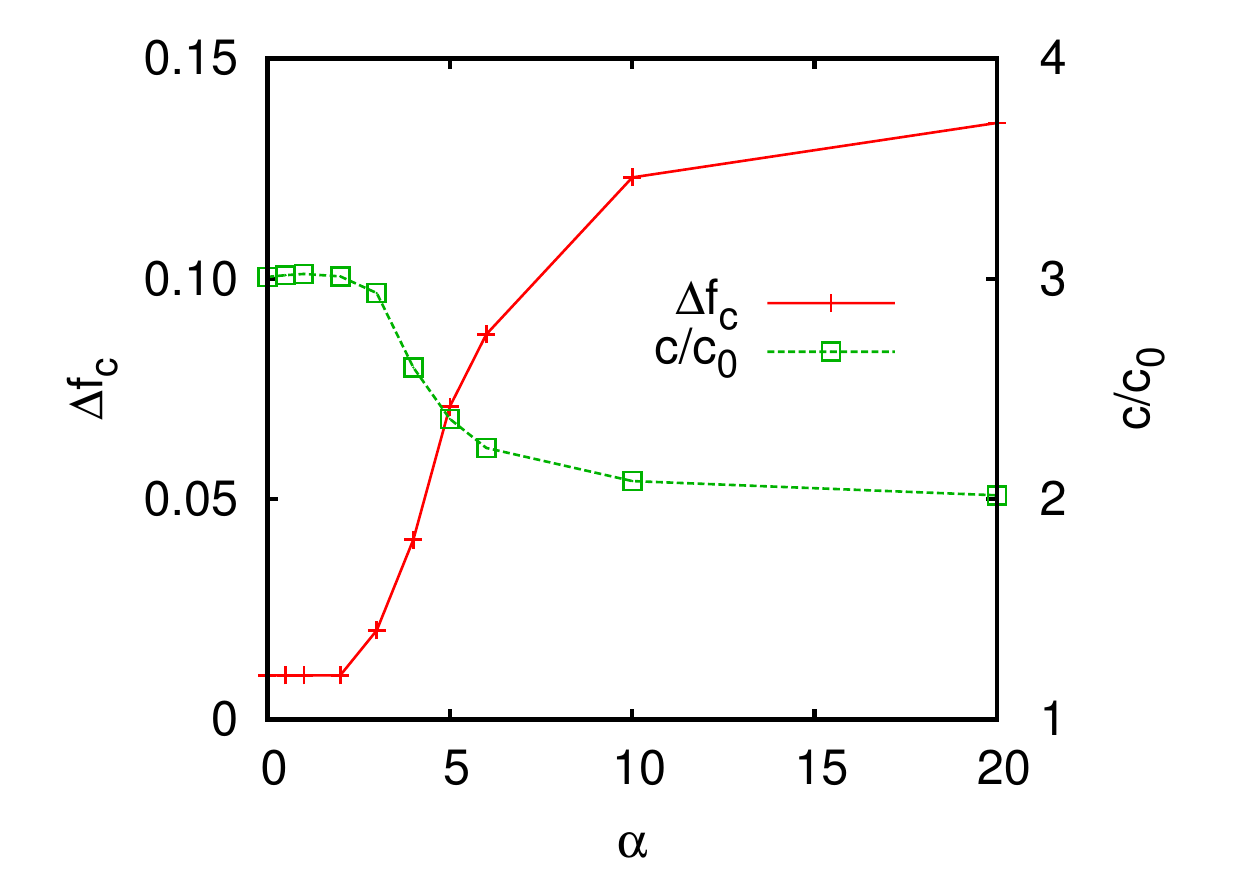}
\caption{\small
  Characteristic quantities for the geographic multilayer WSN model.
  The difference $\Delta f_c$ between the percolation thresholds is shown as a function of $\alpha$. The ratio $c/c_0$ is also shown, where $c$ ($c_0$) is the number of communities a node belongs to for the multilayer (single layer) model. The figure was taken from~\cite{murase2014multilayer}.
}
\label{fig:df_cc_geo}
\end{center}
\end{figure}

The link percolation analysis was conducted for the geographic model using various $\alpha$ values. When $\alpha$ is close to zero, the model does not show the Granovetterian structure, i.e., $\Delta f_c \approx 0$. Since the network in this case has no significant geographic effect, the model is essentially equivalent to the naive multilayer model. As $\alpha$ gets larger, $\Delta f_c$ starts to become larger than zero. The dependence of $f_c$ on $\alpha$ is shown in Fig.~\ref{fig:df_cc_geo}. The difference $\Delta f_c$ becomes larger with increasing $\alpha$ and seems to get saturated around $0.15$. As shown 
in Fig.~\ref{fig:df_cc_geo}, the network for $\alpha=6$ exhibits a Granovetterian structure due to $f_c^a$ and $f_c^d$ being significantly different with $\Delta f_c \approx 0.1$.

A small sample of networks ($N=300$) for $\alpha = 6$ is shown in Fig.~\ref{fig:multilayer_snapshots}(b). The network for a large $\alpha$ clearly shows a community structure. Due to the correlations between the layers, the network has overlapping communities while maintaining the Granovetterian weight-topology relationship.

The geographic extension of the model produces a region of $\alpha$, where a multilayer Granovetterian structure exists. Now we have to check whether our construction would also lead to the enhancement of the overlapping of communities. We have analyzed the aggregate networks by the method of Ahn~\emph{et al.}~\cite{Ahn2010Link} to calculate $c/c_0$. Here $c$ ($c_0$) denotes the average number of communities a node belongs to, for the multilayer model (for the corresponding single-layer model).
If the ratio $c/c_0$ is larger than 1, nodes have significant amount of overlapping communities due to enhancement by the multilayer structure.
Figure~\ref{fig:df_cc_geo} shows the dependence of this quantity on $\alpha$.
The ratio ${c}/{c}_0$ decreases rather rapidly when $\alpha$ increases from 0, and then it reaches the limit value of 2.
This means that for sufficiently large $\alpha$ we have {\it both}
Granovetterian properties and the enhancement in the number of
overlapping communities.

The coexistence of the Granovetterian structure and the enhanced overlapping communities require non-trivial correlations between the layers.
For example, we have tested a model, where the second layer of the network is constructed by replicating the first layer and then the fraction $p$ of the nodes in the second layer gets shuffled. That is, for each pair of nodes $i$ and $j$, with a probability of $p$ all links $ik$ are exchanged with $jk$ only in the second layer.
When $p$ increases from $0$ to $1$, we find a crossover from the single-layer model to the naive multilayer model and $\Delta f_0$ changes from a finite positive value to zero as $p$ approaches $1$. We measured the overlap $c/c_0$ also for this model, however, the overlap starts to increase only when the Granovetterian correlation between link weight and topology is already wiped away. There is no region of $p$, where both required properties can simultaneously be observed. Thus an appropriate introduction of the inter-layer correlation, as shown for the geographic model, is necessary.

\subsection{Modeling channel selection and sampling bias}

In recent years empirical analysis of the society has speeded up due to access to immense amount human-related ICT data~\cite{Onnela2007Structure,Zhao2011Social, Isella2011Whats, Eckmann2004Entropy, Ugander2011Anatomy, Kwak2010What, Onnela2007Analysis}. Most of the data show consistent features as summarized in Subsection~\ref{Sec:stylized}. However, as described in the Introduction, data are usually collected from a single communication channel, i.e., a single-layer of the multiplex depicted in Fig.~\ref{Fig:Fig1}(b). Consequently, the following question remains to be answered: To what extent do results of a single-layer of this multiplex network represent the characteristics of the combined, and thus full social network?

Of course, the best solution would be to combine data from all single
channel layers. This can be done, for example, for transportation
networks~\cite{gallotti2016lost}, but due to technical, privacy, and
legal issues it has been impossible for social data. Therefore, except for
cases of reality mining~\cite{Eagle2009Inferring,
Jo2012Spatiotemporal, Stopczynski2014Measuring} with relatively small
number of participants, we are left with single-layer data resulting
from a non-trivial sampling mechanism that introduces a bias as
compared to the complete, aggregate network, what we are mainly
interested in. In this Subsection we analyze such a sampling by modeling
the channel selection process.

It is known that in order to preserve the original statistics of the network one has to do a careful sampling~\cite{Stumpf2005subnets, Lee2006Statistical} and we cannot expect that the way people select their communication channels will obey these rules.  This is perhaps most apparent in the form of the degree distribution, which was found to be a decreasing function in almost all datasets (see Subsection~\ref{Sec:stylized}). However, it contradicts to all expectations that the most probable case is, when someone has just one single friend. On the contrary, one would rather expect that maximum of the distribution is at a degree of order of the Dunbar number ($\sim150$)~\cite{Dunbar2011Constraints}. The question arises: If the degree distribution is so much distorted by sampling, then to what extent can one trust observations of other properties, e.g., of assortativity, when only a single communication channel is analyzed?

In order to answer this question we devise a simple sampling model
motivated by natural concepts of human communication channel selection
mechanism and show that this way we can reproduce some stylized
facts obtained by empirical ICT data analysis even from 
random networks~\cite{torok2015what}. In particular, we show how the expected peaked degree distribution gets transformed to a monotonic behavior. 

\begin{figure}[t]
\begin{center}
 \includegraphics[width=.90\columnwidth]{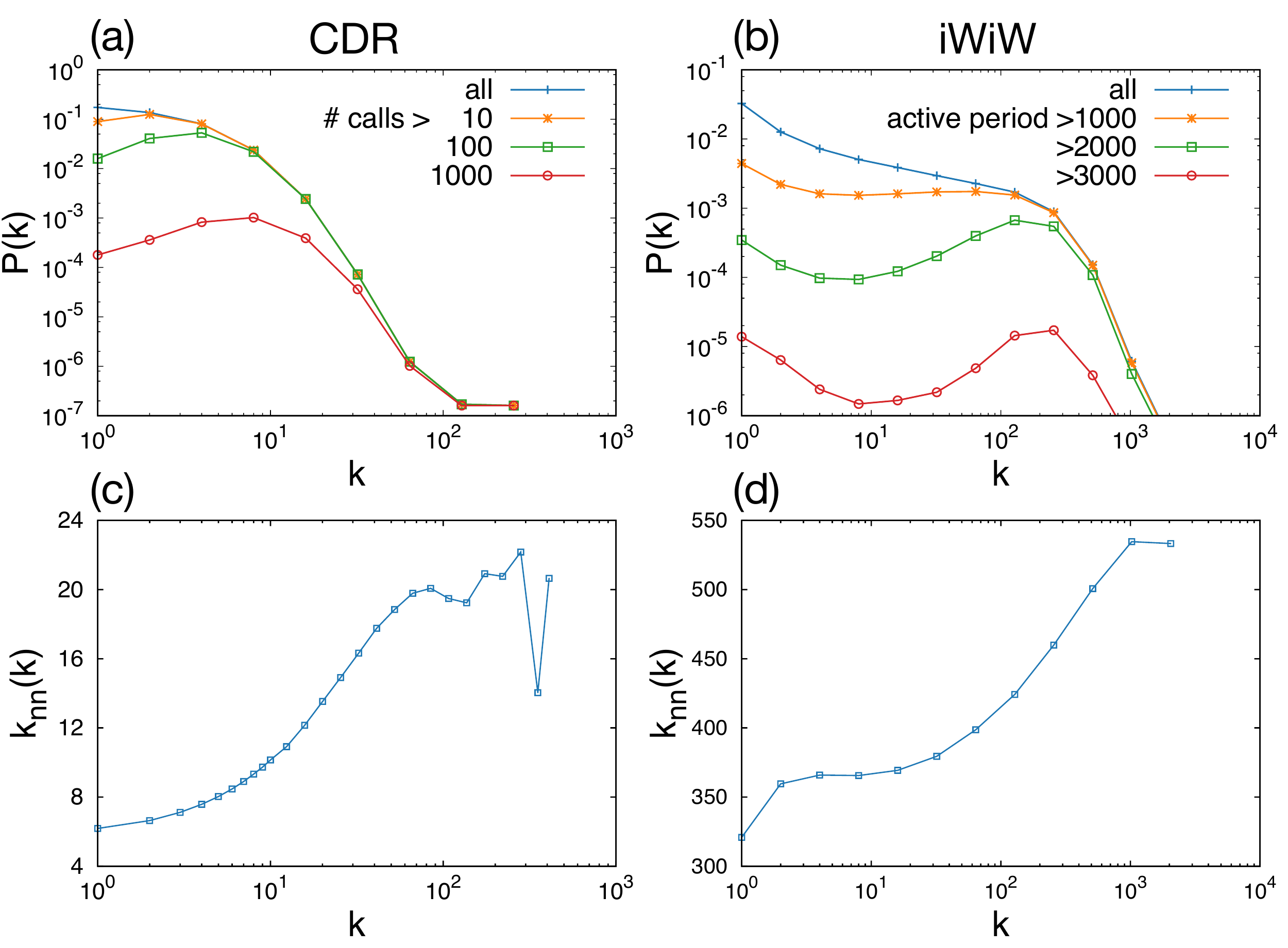}
 \caption{\label{Fig:empiricalsample} \small
     Empirical results of the CDR and the Hungarian online social network iWiW dataset: Degree distributions $P(k)$ of CDR (a) and iWiW (b). We also show $P(k)$-s for nodes with different activity (CDRs) and time spent with the service (iWiW). In the bottom row the average degree $k_{nn}$ of neighbors for nodes with degree $k$ is depicted for the CDR (c) and iWiW (d).
}
\end{center}
\end{figure}

We focus our analysis on two general quantities, namely the degree
distribution $P(k)$ and the average degree of neighbors $\knn(k)$ of nodes with degree $k$. In Fig.~\ref{Fig:empiricalsample} empirical results are shown for two datasets, i.e., a CDR dataset~\cite{karsai2011small} and iWiW dataset, the Hungarian online social network that was closed in 2013 but for 2--3 years it hosted more than two thirds of the population with Internet access in Hungary~\cite{Lengyel2015Geographies}. Both datasets show similar qualitative features even though they are quite different, e.g., the average degree is $7.7$ for CDR and $220$ for iWiW dataset. The degree distributions for all nodes in Fig.~\ref{Fig:empiricalsample}(a) and (b) are monotonically decreasing functions. Interestingly, if we apply a filter and keep only those users who are sufficiently dedicated to the service, meaning large numbers of calls in case of the CDR dataset and longer active periods in the iWiW dataset, the peaked nature of the degree distributions gets brought out. In Fig.~\ref{Fig:empiricalsample}(c) and (d) $k_{\rm nn}$ increases with the degree $k$, indicating assortativity. Even the behavior of the second derivative looks similar. It is, however, unclear whether these features reflect the properties of the underlying social network or they are the results of the sampling bias.

We therefore try to model the process by which people choose communication channels and see how a surrogate network representing the (unknown) true social network is transformed. We have chosen three different networks: Regular random graph (RR), Erd\H{o}s-R\'enyi graph (ER), and link deletion version of the weighted social network (WSN)~\cite{Murase2015Modeling}. All three networks have a peaked degree distribution, and RR and ER show no assortative mixing, but WSN does.

When people want to communicate they have to choose the channel of communication. Naturally people have diverse preferences and may favor different communication channels. However, sticking to someone's favorite does not make sense, e.g., writing a message on a chat server to someone who checks his account only weekly is rather meaningless, and so is calling someone, who never picks up the muted phone. In order to make the communication successful one has to resort to the least uncomfortable channel to both of them. To make it more quantitative we assign an \emph{affinity} to an individual $i$ towards a communication channel $v$ by $f_i^v$. We assume that the probability of choosing the communication channel $v$ for individuals $i$ and $j$ is proportional to the smaller of the two affinities: $p_{ij}=\min(f_i^v,f_j^v,1)$. Thus the probability of a link to exist in the layer $v$ will also be proportional to $p_{ij}$.

Our model for the sampling effect of a single communication channel is thus defined as follows: Let us consider a surrogate network. Each node is given a randomly chosen affinity $f$ towards this specific communication channel. The affinities are taken from an exponential distribution, reflecting that there are always more people who put small effort in a specific ICT service and there are few who are really addicted to it:
\begin{equation}
P(f)=\frac{1}{f_0}e^{-f/f_0}.
\label{Eq:affinity}
\end{equation}
The links in the sampled network are kept with probability
\begin{equation}
p_{ij}=\min(f_i,f_j,1).
\label{Eq:selection}
\end{equation}
All nodes which have at least one link are kept for the sampled
network.

\begin{figure}[t]
\begin{center}
 \includegraphics[width=.9\columnwidth]{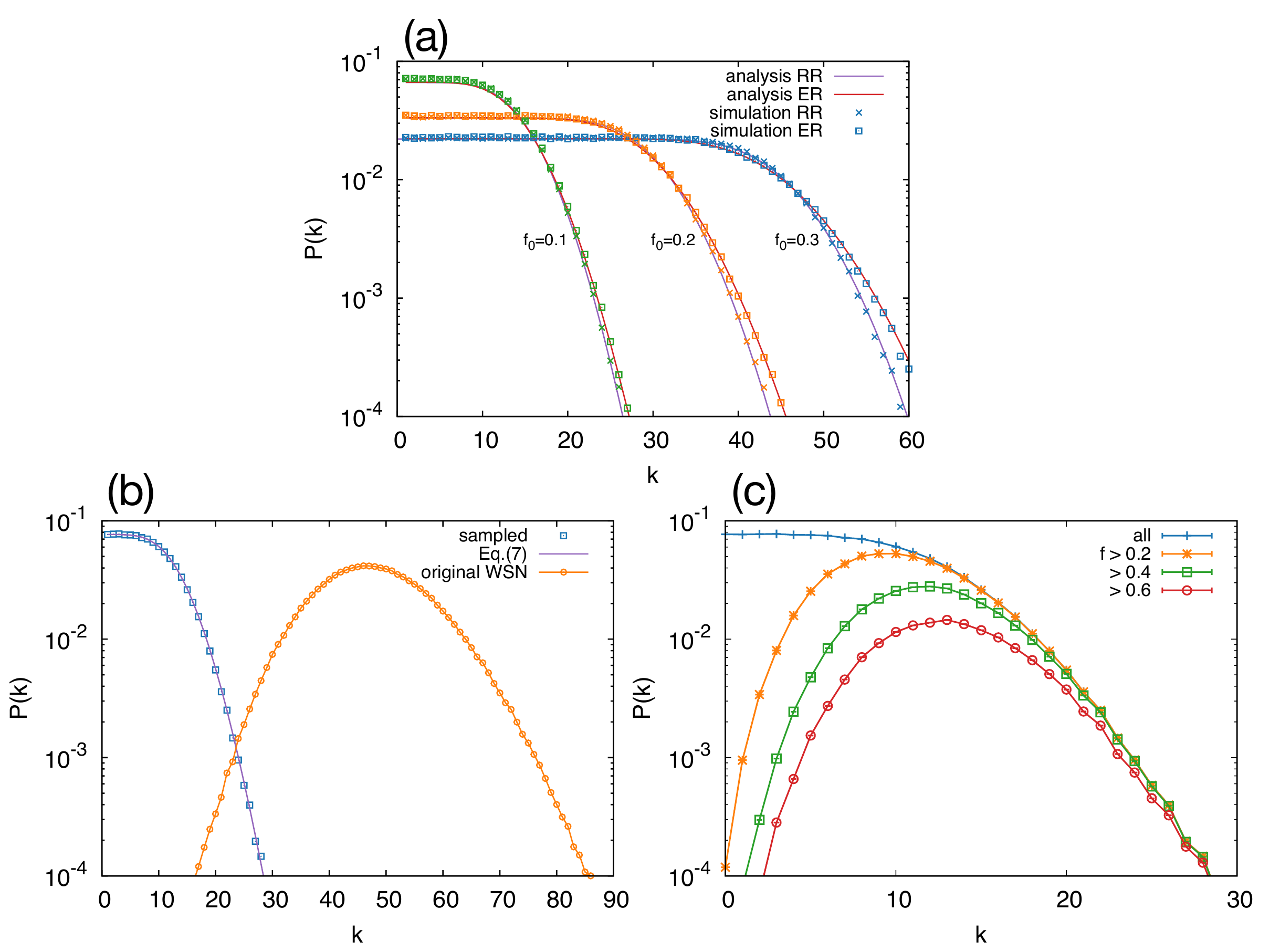}
 \caption{\label{Fig:pkresult} \small
  (a) Degree distributions of sampled networks when RR and ER with $k_0=\langle k\rangle =150$ and $N=10^4$ are used as surrogate networks, using $f_0=0.1$, $0.2$, and $0.3$. Solid lines denote analytic solutions of Eq.~(\ref{Eq:pkanal}). (b) Degree distributions of original and sampled networks using the WSN model as surrogate network with $f_0=0.3$. The solid line denotes the degree distribution obtained using Eq.~(\ref{Eq:pkanal}). (c) Degree distribution of the sampled network from WSN using $f_0=0.3$ and those when restricted only for nodes having an affinity above the indicated threshold.
}
\end{center}
\end{figure}

We have tested our sampling model for the following surrogate networks: ER and RR with average degree of $k_0=\langle k \rangle=150$ and WSN with $\langle k \rangle\simeq 47.8$\footnote{Here we used Link-Deletion WSN model proposed in \cite{Murase2015Modeling}. The parameters to generate WSN are $N = 10^4$, $p_{\Delta} = 0.07$, $p_r = 0.0007$, $p_{ld} = 0.0015$, and the maximum time step $t_{max}=50000$.} The results are shown in Figs.~\ref{Fig:pkresult}(a) and (b). Clearly, the originally peaked distribution has become monotonically decreasing by sampling. It is also interesting that the shape of the curve depends only very little on the original degree distribution, as demonstratively shown in Fig.~\ref{Fig:pkresult}(a). Here we find the marginal difference between the degree distributions of RR and ER.

We can carry out a similar filtering as before for the single-layer empirical data by selecting nodes dedicated to the channel. The high affinity nodes show progressively peaked degree distributions in Fig.~\ref{Fig:pkresult}(c) which indicates that indeed the properties of such nodes are closer to the original network than low affinity ones.

The sampled degree distribution can be calculated analytically~\cite{torok2015what} using the fact that the affinities are assigned randomly so there is no correlation between the affinities of neighboring nodes in the surrogate networks. The degree distribution $Q_{k_0}(k)$ for the RR network with degree $k_0$ is

\begin{equation}
Q_{k_0}(k)=
\frac{1}{f_0(k_0+1)}I_{\left(\frac{f_0}{1-f_0}\right)}(k+1,k_0-k+1),
\end{equation}
where $I_x(a,b)$ denotes the regularized beta function. 

For general degree distributions, but yet for the case of uncorrelated affinities the degree distribution can be obtained as a weighted sum:
\begin{equation}\label{Eq:pkanal}
P(k)=\sum_{k'=0}^\infty P_0(k')Q_{k'}(k).
\end{equation}
Equation (\ref{Eq:pkanal}) is verified against the numerical data in
Figs.~\ref{Fig:pkresult}(a) and (b) and the match is perfect.

\begin{figure}[t]
\begin{center}
 \includegraphics[width=.9\columnwidth]{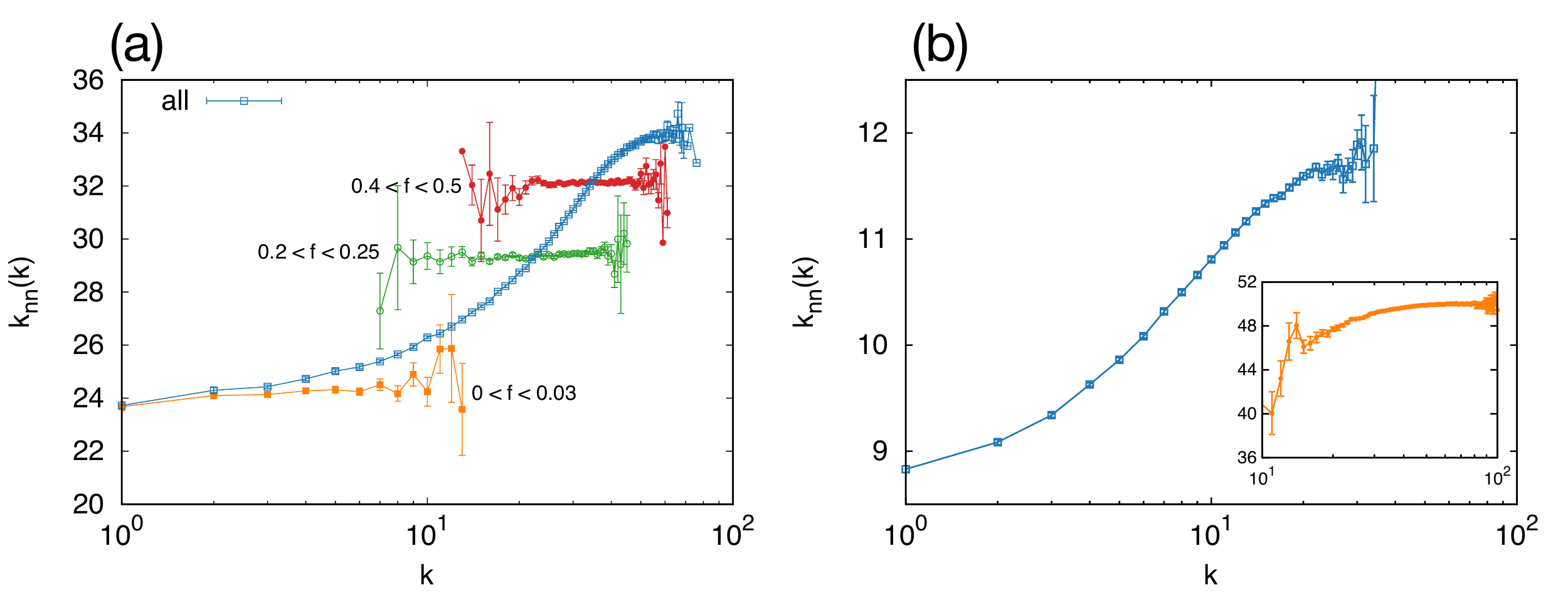}
 \caption{\label{Fig:sampleknn} \small
 (a) Average degrees of neighboring nodes as a function of the node degree for sampled network using Erd\H{o}s-R\'enyi graphs as surrogate networks for all nodes, and for a range of node affinity. (b) Assortativity of the sampled network as a function of the degree when WSN is used as a surrogate network.  The inset shows the case for the surrogate network.
}
\end{center}
\end{figure}

We have shown that our channel selection model reproduces the observed effect, namely the transformation from a peaked degree distribution to a monotonically decreasing distribution due to sampling. Now we turn our attention to the assortativity as calculated from $\knn(k)$, the average degree of the neighbors for nodes with degree $k$, see Fig.~\ref{Fig:sampleknn}. The sampled results both for ER and WSN cases show similar assortative mixing as the empirical data, even though it is known that ER has degree independent $k_{\rm nn}$. This demonstrates that sampled networks can show similar assortative behavior irrespective of their original properties. Again, considering nodes in each given affinity range, we find flat behavior for $\knn(k)$ for those nodes as in the surrogate networks, as 
depicted in Fig.~\ref{Fig:sampleknn}(a).

A remark has to be made at this point. Our channel selection model as described by Eqs.~(\ref{Eq:affinity}) and~(\ref{Eq:selection}) is certainly a crude approximation of reality. In order to check the robustness of our results, we applied the generalized mean instead of taking the minimum of the affinities for the selection rule: 
\begin{equation}
p_{ij}=\Big(\frac{f_i^\beta+f_j^\beta}{2}\Big)^{1/\beta},
\end{equation}
with $\beta \to -\infty$ providing the rule of Eq.~(\ref{Eq:selection}) used above. We have shown that we have a decreasing degree distribution in the sampled networks only when $\beta$ is negative and that assortativity is generated for this parameter region even if we use uncorrelated surrogate networks~\cite{torok2015what}.

Our simple
model of communication channel selection shows that the sampled network resulting
from this selection mechanism may seriously distort the properties of the original network. As most of the nodes have small affinity,
their social network will be poorly represented in a given ICT network.
The nodes having high degree in the sampled network are not
necessarily the ones that had most contacts in the original network
but the ones with high affinity towards this particular service. This
distorts the network in such a way that new features can be observed.
The sampling model presented here has so strong influence on the
network properties that it may completely hide the original ones and
shows the biased properties instead. This emphasizes that 
sngle-channel empirical data should be handled with care.

Our study also demonstrated that we may get some insight into the real structure of the original network properties if the analysis is restricted to a subset of well embedded nodes from the sampled network. In our calculations we used the affinity of the nodes as a measure of this embeddedness but our results on CDRs and SNS data indicate that activity or time spent with the service may also be used for this purpose.

\subsection{Summary and discussion}

In this Chapter we have discussed that the society can be considered as a multiplex with respect to the nature of the links reflecting the contexts of the interactions between the persons (generative aspect) or from the point of view of the communication channel (data collection aspect), see Fig.~\ref{Fig:Fig1}.
We have shown how the Granovetterian structure and the overlapping communities can be maintained in a multiplex model. In  order to do so, we started from the single-layer WSN model~\cite{Kumpula2007Emergence} and generalized it to a multiplex. However, a naive introduction of multiple layers to single-layer WSN models breaks the Granovetter-type weight-topology
relation, so instead we introduced geographic correlations between the layers. 

Our results have several implications. Firstly, we have shown the importance of correlations between layers. Moreover, it seems that specifically geographic correlations may play a key role in maintaining the stylized facts in a multiplex weighted network. It is worth noting that the peculiar role of geographic correlations was observed earlier~\cite{Kosmidis2008Structural, Barthelemy2011Spatial}, and 
for interdependent networks~\cite{Li2012Cascading}. We mention that communities may organize themselves along various diverse but common attributes like sharing working places, classes at universities, joint interest, e.g., in sport, and residential districts. However, all these have some geographic aspect. In fact, even in the digital era distance is not ``dead''~\cite{Goldenberg2009Distance, Lengyel2015Geographies} contrary 
to some earlier speculations~\cite{Cairncross2001Death}. Of course, the consideration of further realistic correlations should improve the model. 

The other multiplex aspect of the society is related to the different communication channels. Due to the fact that our data analytics mostly relies on observations from a single communication/interaction channel the question arizes:  To what extent ICT data can tell us about the structure of the entire social network of people, as all such data are incomplete and capture only a part of the whole plethora of social relationships. This is closely related to the important question of channel selection, which we have attempted to model here.

While ICT services are diverse, we nevertheless observe some common features, e.g., that they all display an overall decreasing degree distribution, which cannot be true for the entire social network and hence should be attributed to the sampling. To investigate the effect of sampling by single channel selection we have modelled how people are using ICT communication services. Using simple assumptions we were able to reproduce robustly the stylized facts of the ICT data, namely the decreasing degree distributions and assortative mixing, even when they were absent in the original surrogate networks. Our results firstly resolve the long lasting contradiction between the observed and expected shapes of the degree distributions. Moreover, they call the attention to the danger of misinterpreting observations from single channel data for the entire social interaction network. At the same time we have also shown that there is a subset of users with high activity, i.e., users who put much effort into the given ICT service, whose characteristics are at least qualitatively in accordance with those of the original surrogate network. This feature hints towards a possible resolution of the problem of the sampling bias. 

Our results rely on the model of channel selection as expressed by Eqs.~(\ref{Eq:affinity}) and~(\ref{Eq:selection}) and their generalizations. We have shown that there is a class of rules that result in the universally observed single channel properties of monotonic degree distribution and assortative mixing. Such class of rules are similar to the minimum rule~(\ref{Eq:selection}), i.e., a person does not select a communication channel with a friend who does not like that channel even if that is the person's favorite.

It should be mentioned at this point that we consider our channel selection model as a first step only in this very interesting problem. Clearly, the assumption of uncorrelated affinities should be revized. Homophily, one of the most important factors in tie formation~\cite{McPherson2001Birds}, implies that there are strong similarities in the affinities of neighbors. Also node properties, like age and gender, should influence affinity values. These features may generate higher order correlations enabling to deal with the effect of sampling on clustering and communities. 

The endeavor of large scale modeling of the society has just started. The activity is increasing and several attempts have already been published. Here we focused on our own contributions but we could mention, e.g., the recent model of Battiston~\emph{et al.} on multi-layer modeling of a given community structure~\cite{Battiston2016Emergence} or the very interesting model of virtual multilayer society  by Klimek~\emph{et al.}~\cite{Klimek2016Dynamical}. Although the models are strong simplifications of the society, we believe that they contribute to the understanding of social structures. Moreover, adequate models enable us to investigate the impact of the structure on dynamic phenomena, e.g., spreading. Future work in such direction is also expected.

{\bf Acknowledgements:} J. K. acknowledges support from EU Grant No. FP7 317532 (MULTIPLEX). J. T. thanks for financial support of Aalto AScI internship programme. Y. M. appreciates hospitality at Aalto University and acknowledges support from CREST, JST.  H.-H. J. acknowledges financial support by Basic Science Research Program through the National Research Foundation of Korea (NRF) grant funded by the Ministry of Education (2015R1D1A1A01058958) and the  framework of international cooperation program managed by the National Research Foundation of Korea (NRF-2016K2A9A2A08003695). This project was partly supported by JSPS and NRF under the Japan-Korea Scientific Cooperation Program. Partial support by OTKA, K112713 is also acknowledged. The systematic simulations in this study were assisted by OACIS~\cite{Murase2014Tool}.  K.K. acknowledges support from Academy of Finland's COSDYN project (No. 276439) and EU’s Horizon 2020 FET Open RIA 662725 project IBSEN.

\bibliography{Multiplex_book_Kertesz_et_al}

\end{document}